# Parallel Vectorized Algebraic AES in MATLAB for Rapid Prototyping of Encrypted Sensor Processing Algorithms and Database Analytics

Jeremy Kepner, Vijay Gadepally, Braden Hancock, Peter Michaleas, Elizabeth Michel, Mayank Varia
MIT Lincoln Laboratory, Lexington, MA, U.S.A.

*Abstract—* The increasing use of networked sensor systems and networked databases has led to an increased interest in incorporating encryption directly into sensor algorithms and database analytics. MATLAB is the dominant tool for rapid prototyping of sensor algorithms and has extensive database analytics capabilities. The advent of high level and high performance Galois Field mathematical environments allows encryption algorithms to be expressed succinctly and efficiently. This work leverages the Galois Field primitives found the MATLAB Communication Toolbox to implement a mode of the Advanced Encrypted Standard (AES) based on first principals mathematics. The resulting implementation requires 100x less code than standard AES implementations and delivers speed that is effective for many design purposes. The parallel version achieves speed comparable to native OpenSSL on a single node and is sufficient for real-time prototyping of many sensor processing algorithms and database analytics.

## I. INTRODUCTION

Modern buildings [Hubbell 2012] and vehicles [Gadepally 2014] contain many networked sensors to monitor and control their environmental, security, and safety systems. The decreasing size and increasing capability of these networked sensors has led to their increased usage on people [Starner 2014]. These wearable technologies provide a wide variety of data that will extend and improve human lives. This trend towards increased use of networked sensors is often referred to as the Internet-of-Things (IoT) [Atzori 2014].

IoT data is stored and analyzed in networked databases accessible to a wide range of stakeholders: users, family members, care providers, operators, manufacturers, and regulators. Sensor algorithms and database analytics are the primary mechanisms for transforming raw IoT data into useful information. Incorporating encryption directly into sensor algorithms and database analytics is one method for adding security protections to IoT systems.

The critical role that IoT data plays in controlling key aspects of society necessitates that a variety of protection measures be included in the design of IoT systems. Figure 1 depicts a common architecture for connecting IoT data sources with users via a range of technologies.

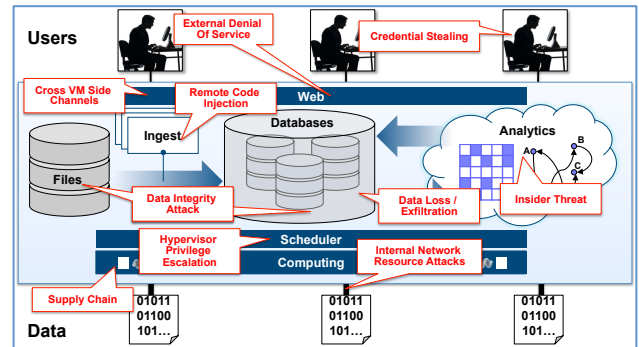
(a) IoT Security challenges

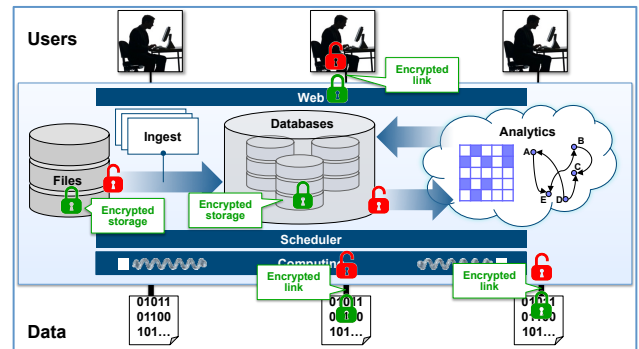
(b) IoT Security current approaches

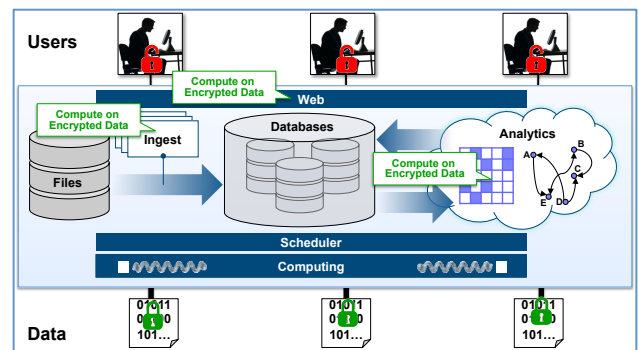
(c) IoT Security vision

Figure 1. Standard architecture for connecting diverse data and users. Shown are big data veracity challenges (a), current approaches (b), and longer term vision (c).

There are many security challenges in such an IoT data processing system (see Figure 1a): external denial of service, credential stealing, cross virtual machine (VM) side channels, VM hypervisor privilege escalation, remote code injection, data integrity attacks, data loss/exfiltration, insider threats, internal network resource attacks, and supply chain attacks [Evans 2013].

These attacks threaten to degrade the availability of the IoT processing system, compromise the confidentiality of the data and analytics used, and violate the integrity of both the original data and the analytic results. There are several approaches to mitigating these security challenges. Data centric protections are particularly useful for preserving the confidentiality of the data. Typical defenses of this type include (see Figure 1b): encrypting the links between users and the IoT processing system, encrypting the links between the data sources and the IoT processing system, encrypting the data in the file system, and encrypting data in the database when it is at rest. These approaches are all significant steps forward in improving the security of an IoT processing system. However, all of these approaches require that the data be decrypted for it to be used inside the IoT processing system, which requires that the keys to the data be available to the IoT processing system thus exposing the IoT processing to any attacker able to breach the boundaries of the system.

One vision (Figure 1c) for an IoT processing system is to have data sources encrypt data prior to transmitting it to the system, have the IoT processing system operate on the data in encrypted form, and only allow authorized users the keys to decrypt the answer for their specific result. Such a system makes the underlying IoT processing technologies oblivious to the details of the data and would go a long way towards mitigating the IoT processing security challenges described above. As a result, the data and processing can be outsourced to an untrusted cloud while preserving the confidentiality of the data and results.

There are several cryptographic tools that one could use to build a system like the one shown in Figure 1c. Fully homomorphic encryption allows for arbitrary analytic computations to be performed on encrypted data without decrypting it while preserving its semantic security (i.e., no information about the data is leaked other than its length). Fully homomorphic encryption has been an active topic of research since its discovery [Gentry 2009]. Nevertheless, the best currently available schemes [Perl 2011, Halevi 2014] have an overhead of $10^5$ or more, making them too slow for use in many systems [Varia et al 2015a].

If one is willing to allow a small amount of information about the encrypted data to be revealed, a higher speed approach is to design protocols that leverage more traditional cryptographic techniques to carry out queries on encrypted data. One example of such a protocol is CryptDB [Popa 2011], which constructs a practical database system capable of handling most types of SQL queries on encrypted data. It uses deterministic encryption, which always encrypts the same data to the same ciphertext, to enable equality queries; order-preserving encryption, which encrypts data in a way that preserves the original order of the data, to enable range queries; and additively homomorphic encryption, which enables summing values directly on encrypted data, to perform basic analytics. Several other protocols achieving alternative trade-offs between leakage and efficiency have been proposed by [Raykova 2012, Pal 2012, Cash 2013, Varia et al 2015b]. Additional solutions are also possible using techniques for secure multi-party computation [Yao 1982, Ben-Or 1988] but these require further improvement to achieve the required speed.

Incorporation of the aforementioned cryptographic techniques into secure algorithms and analytics requires a co-design with encrypted computation and encrypted query. Currently these areas have few common design processes or tools and employ different design criteria, mathematics, and programming.

A typical sensor algorithm involves filtering the raw data, correlating with known signatures, detecting significant correlations, and summarizing the precise location of the detections in the data. Throughout the sensor algorithm development process the designer seeks to balance a variety of system characteristics such as signal-to-noise ratio, probability of detection, and probability of false alarm. The primary mathematics of sensor algorithms is linear algebra, and the most common algorithm development environment is MATLAB.

A typical database analytic might first select records from a database, perform a database join on the records, construct a graph out of the records, and cluster the graph to find records of most interest. During the design of a database analytic, the designer will simultaneously balance database system characteristics such as ingest rate and query latency. The primary mathematics used is set theory, and SQL is the de-facto language of database environments.

Encrypted computation utilizes a variety of techniques such homomorphic computation, fully homomorphic computation, multi-party computation, and computing on masked data. The design of encrypted computation focused on developing adversary models and understanding the information leakage of various encryption techniques with respect to different adversary models. The mathematics of encryption relies on number theory and in particular Galois Fields. There are a number of environments for developing encrypted computation, and Python is widely used.

Encrypted query relies on various approaches for obscuring the data while still allowing some search to occur: for example, deterministic and order preserving encryption. These are often layered in an "onion" so that the strongest encryption is on the outside of the data and can be "peeled" to the weaker levels as needed. Furthermore, the actual patterns of query can also be designed in a privacy-preserving manner so that the query itself reveals as little information as is possible. Encrypted query is similar to encrypted computation and focuses on minimizing information leakage with respect to specific adversary models. Likewise similar Galois Field mathematics are also used. The dominant programming environment is the same SQL used for standard database analytics.

An important step towards enabling the co-design of secure algorithms and analytics is to provide common environments where co-design can occur. This requires both mathematical and technological unification of the environments. Associative array provides a common mathematics for sensor algorithms and database analytics [Kepner 2012, Chaidez 2014, Gadepally 2015a, Kepner 2015, Kepner & Jansen 2016]. Computing on Masked Data has demonstrated encrypted computation and query on associative arrays using external encryption libraries [Kepner 2014, Gadepally 2015b]. The next step is to integrate encryption mathematics (Galois Fields) directly into these associative array environments. The advent of high performance Galois Field mathematics in high level programming environments such as MATLAB and Magma [Agullo 2009] now make it feasible to incorporate encryption directly into sensor processing and database analytic design environments. This paper demonstrates this capability with the Advanced Encryption Standard (AES) algorithm and shows that fast encryption can be performed in a co-design environment suitable for sensor algorithm development and database analytic design.

We caution here that MATLAB is primarily an algorithm prototyping tool. As such, our implementation is not intended directly for use in practice. Any production implementation of AES should take into account side channel attacks that can exploit variations in the software's running time [Kocher 1996] or power consumed [Kocher 1999, Chari 2002, Peeters 2007] in order to learn the secret key. Such attacks are very effective in practice, often requiring just tens or hundreds of measurements to learn the key completely [DPA contest, Osvik 2006]. Production implementations of AES require careful handling to provide proper defenses against side channel attacks [Messerges 2001].

The outline of the rest of this paper is as follows. Section II gives a brief introduction to the mathematics of AES. Section III describes the implementation of AES using the Galois Field primitives found in the MATLAB Communications Toolbox. Section IV presents the comparative results of this implementation versus other implementations. Section V presents our conclusions and plans for future work.

## II. ADVANCED ENCRYPTION STANDARD

AES is a specification established by U.S. National Institute of Standards and Technology (NIST) in 2001. AES is based on on the Rijndael algorithm developed by Joan Daemen and Vincent Rijmen [Daemen & Rijmen 2001]. Rijndael is a family of ciphers with different key sizes, block sizes, and modes. The standard blocksize for AES is 128 bits (8 bytes), which means that unencrypted data (plaintext) is transformed into encrypted data (ciphertext) into blocks 8 bytes at a time. The encryption can be performed with keys of varying lengths: 128 bit, 192 bit, or 256 bit. For this work, the focus will be on 128 bit encryption. Finally, the encryption scheme can be utilized in a variety of modes. Electronic Codebook (ECB) mode encrypts each 8 byte block of plaintext independently of all of the others. Cipher Block Chaining (CBC) mode combines the plaintext with an Initialization Vector (IV) and then encrypts the data. The ciphertext of the 8 byte block is then used as the IV for the next 8 byte block. For this work, CBC mode is implemented.

The core mathematics of AES encryption is a Galois Field arithmetic [Artin 2011], where the bits in each byte represent the coefficients of a polynomial. The decimal, binary, and Galois Field representations of various numbers are as follows:

$1 = 00000001 = 1$
$3 = 00000011 = x + 1$
$7 = 00000111 = x^2 + x + 1$
$15 = 00001111 = x^3 + x^2 + x + 1$
$31 = 00011111 = x^4 + x^3 + x^2 + x + 1$
$63 = 00111111 = x^5 + x^4 + x^3 + x^2 + x + 1$
$127 = 01111111 = x^6 + x^5 + x^4 + x^3 + x^2 + x + 1$
$255 = 11111111 = x^7 + x^6 + x^5 + x^4 + x^3 + x^2 + x + 1$

Addition and multiplication of numbers in Galois Field are performed using the normal rules of polynomial addition and multiplication of binary coefficients

$1 + 3 = 00000001 + 00000011$
$\quad = 1 + (x + 1)$
$\quad = x + 2$
$\quad = x + 0$
$\quad = 00000010$
$\quad = 2$

In some cases this would result in a higher order polynomial that cannot be represented by 8 bits

$(x^7)(x^5) = x^{12}$

To solve this problem, the results of all calculations are computed modulo the AES polynomial

$100011011 = x^8 + x^4 + x^3 + x + 1$

Hence

$(x^7)(x^5) \mod (x^8 + x^4 + x^3 + x + 1) = x^7 + x^5 + x^3 + x + 1$

Letting addition, multiplication, and matrix multiplication be defined by the aforementioned Galois Field arithmetic, the AES encryption algorithm can be written succinctly as follows

**P**, **C** : $\mathbb{GF}_{8b}^{N \times M}$    *Matrices of input/output in 8bit Galois Field*
**V** : $\mathbb{GF}_{8b}^{N \times 16}$    *Matrix of IVs in 8bit Galois Field*
**K** : $\mathbb{GF}_{8b}^{R \times N \times 16}$    *Tensor of key schedules in 8bit Galois Field*
**s** : $\mathbb{GF}_{8b}^{256}$    *s-box in 8bit Galois Field*
**M** : $\mathbb{GF}_{8b}^{4 \times 4}$    *Mix matrix in 8bit Galois Field*
**i** = [1:16]    *Set of 16 rows*
for each block of **i** rows
   **C**(:,**i**) = **P**(:,**i**) + **V** + **K**(1,:,:)    *Add IV and round key*
   for each round r
     **C**(:,**i**) = **s**( **C**(:,**i**) + 1)    *Apply s-box*
     **C**(:,**i**$_1$) = **C**(:,**i**$_2$)    *Shift rows*
     **C**(:,1:4) = (**M** **C**(:,1:4)$^T$)$^T$    *Multiply by M*
     **C**(:,5:8) = (**M** **C**(:,5:8)$^T$)$^T$    *Multiply by M*
     **C**(:,9:12) = (**M** **C**(:,9:12)$^T$)$^T$    *Multiply by M*
     **C**(:,13:16) = (**M** **C**(:,13:16)$^T$)$^T$    *Multiply by M*
     **C**(:,**i**) = **C**(:,**i**) + **K**(r,:,:)    *Add round key*
   **V** = **C**(:,**i**)    *Copy back to IV*
   **i** = **i** + 16    *Increment rows*

where $^T$ denotes the transpose of the matrix, and for 128 bit AES encryption there are 10 rounds. The above algorithm will encrypt N plaintexts of length M simultaneously with the same key. The most computationally expensive part of the algorithm is the four mix matrix multiplications. Encrypting many plaintexts at the same time allows these operations to be performed on all the messages using efficient matrix multiplication operations. This technique of combining many items together so they can be performed simultaneously is often referred to as *vectorization* [Birkbeck 2007].

AES decryption can be described in much the same way by reversing the steps of AES encryption. Likewise, the IVs, key schedules, and s-box can also be calculated using simple algorithms.

## III. MATLAB IMPLEMENTATION

The MATLAB implementation of AES has three main goals. The first goal is to make the cryptographic technique easy to understand. The second goal is to provide relatively fast encryption/decryption of many short messages of the type that would be used in a sensor algorithm or a database analytic. The third goal is to provide reasonable speed so that a sensor algorithm or database analytic can be quickly tested on sufficiently large data to determine its effectiveness.

Easy-to-understand implementations allow sensor algorithms, database analytics, and cryptographic techniques to be combined in a way that they can be adapted to a variety of applications. In a MATLAB context, easy-to-understand means that there is a strong correspondence between the MATLAB code and the corresponding mathematics. To meet this goal, the high level Galois Field mathematics found in the MATLAB Communication Toolbox is used that allows the mathematics of AES to be implemented in manner that is very similar to the written mathematics.

Sensor algorithms and database analytics perform their operations on many relatively small data elements such as a sensor measurements or a short strings of characters. Thus cryptographic techniques applied to this data should be optimized for many short messages. Many standard cryptographic implementations are implemented to run well on long messages. Vectorizing cryptographic algorithms so that they perform on many messages at once allows the overall speed of the algorithm on short messages to be increased.

The design process of sensor algorithms and database analytics is an iterative, trial-and-error process whereby the designer implements mathematical techniques to achieve their goals. Throughout this process the implementations are tested on ever increasing datasets to validate the mathematics. Extremely slow implementations can be a significant impediment to the design process and requires the implementation must have good overall speed. A simple way to deliver good overall speed is to employ parallel computing. Fortunately, many vectorized implementations are relatively easy to run in parallel. In this case, since the encryption and decryption of each message is independent, this provides a natural approach to running the algorithm in parallel. Likewise, parallel MATLAB [Kepner 2009] provides the necessary software infrastructure for easily running a MATLAB program on a parallel computer.

Employing the aforementioned techniques, a MATLAB implementation of the AES encryption algorithm can be written succinctly as shown in Appendix A.

## IV. COMPARATIVE RESULTS

The MATLAB implementation listed in Appendix A is compared to three other AES implementations. The two mostly widely recognized AES MATLAB implementations ([Bucholz 2001] and [Matejka 2011]) are very similar to each other. They both focus on providing an AES implementation in MATLAB primarily for illustrative purposes. Neither of these implementations uses the Galois Field primitives found in the Communications Toolbox. The third implementation is a MATLAB binding to the Open Secure Socket Layer [OpenSSL] implementation of AES written in C [Gadepally 2015]. OpenSSL is among the most widely used encryption libraries.

The three goals of the MATLAB Communication Toolbox implementation were for it to be easy to understand, have good speed on many short messages, and have good overall speed. Easy to understand in this context refers to the correspondence of the implementation to the mathematics. Visual inspection of mathematics and the code shows a strong correspondence. A proxy for the ease of understanding is the relative volume of the code measured in software lines of code (SLOC) using a code counting tool [SCLC]. Figure 2 shows the code volume for all four implementation of AES in MATLAB. The Communications Toolbox implementation is 3x smaller than the other MATLAB implementations, 10x smaller than the MATLAB binding to OpenSSL, and 100x smaller if the underlying OpenSSL C code is included.

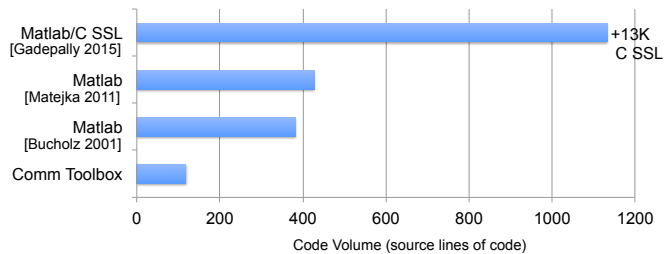

Figure 2. Code volume of different implementations of AES in MATLAB. The Communications Toolbox implementation is 3x smaller than the other MATLAB implementaions, 10x smaller than MATLAB binding to OpenSSL, and 100x smaller if the underlying OpenSSL C code is included.

The speed of the Communications Toolbox implemenation for messages of different size and different number of messages are shown in Figure 3. The top curves show the rate for 16 byte messages with varying numbers of messages. The bottom curves show the rate for 128 messages with varying message lengths. As expected, the vectorized Communications Toolbox implemenation peforms well for many short messages.

Figure 4 is the same as Figure 3 but also includes the rate of the other three implementations on many 16 byte messages. As expected, the OpenSSL implementation is 10x faster. The other MATLAB implemenations are 100x slower. Thus, the Communications Toolbox implementation achieves nearly 10% of the rate of a highly optimized C library with only 1% of the code.

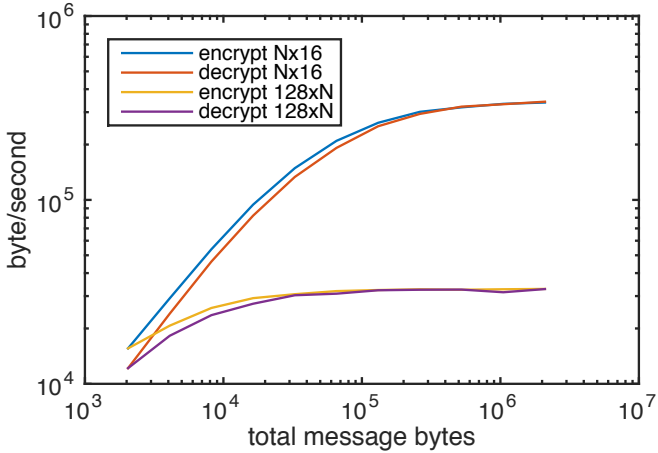

Figure 3. Encryption and decryption rates of the MATLAB Communications Toolbox implemenation as a function of total messsage size. The top curves show rate for 16 byte messages with varying numbers of messages. The bottom curves show the rate for 128 messages with varying message lengths. As expected, the vectorized Communications Toolbox implemenation peforms well for many short messages.

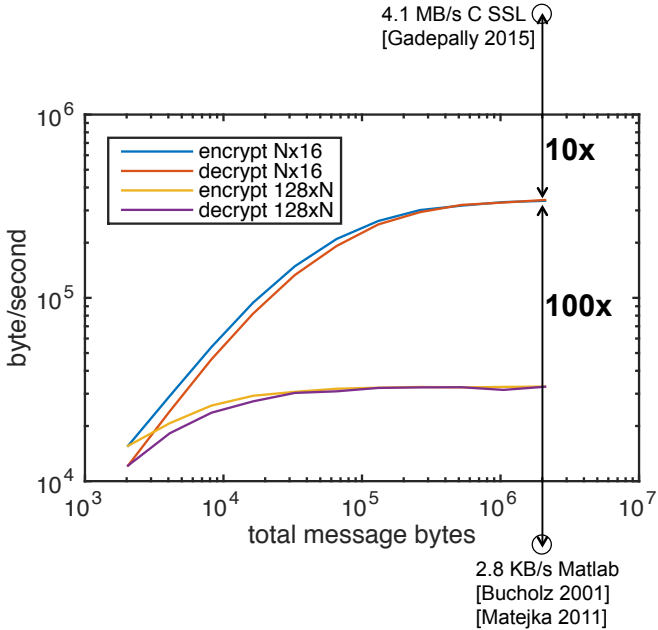

Figure 4. Comparison of encryption and decryption rates of the MATLAB Communications Toolbox implemenation with other implementations. OpenSSL implementation is 10x faster. The other MATLAB implementations are 100x slower.

The overall speed of the Communications Toolbox implementation can be improved by running it in parallel. Because this implementation is designed to run well on many small messages, this provides an easy mechanism for running the code in parallel. Different groups of messages can be processed independently on different processors. Figure 5 shows that a speedup of 20x was achieved using 24 processor cores on a single node of a compute cluster.

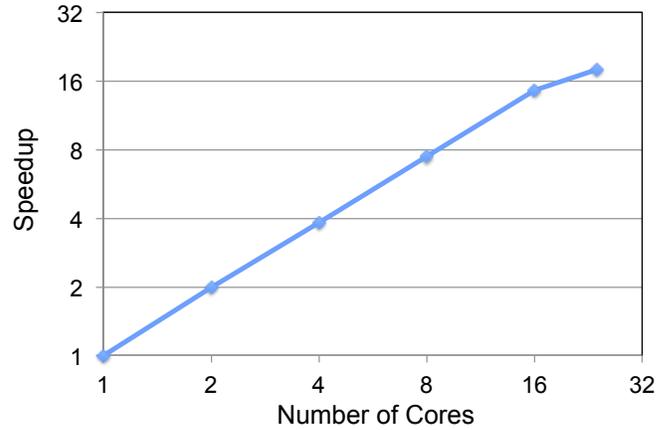

Figure 5. Speedup versus number of cores used for MATLAB Communication Toolbox implementaion. A speedup of 20x was achieved using 24 processor cores on a single node of a compute cluster.

## V. CONCLUSIONS & FUTURE WORK

The increasing use of networked sensor systems and networked databases has led to an increased interest in incorporating encryption directly into sensor processing and database algorithms. MATLAB is the dominant tool for rapid prototyping of sensor processing algorithms and has extensive database analytics capabilities. The advent of high level and high performance Galois Field mathematical environments allows encryption algorithms to be expressed succinctly and efficiently. This work leverages the Galois Field primitives found in the MATLAB Communication Toolbox to implement a mode of the Advanced Encrypted Standard (AES) based on first principals mathematics. The resulting implementation requires 100x less code than standard AES implementations and delivers speed that is effective for most algorithm development purposes. The parallel version achieves speed comparable to OpenSSL on a single node and is sufficient for real-time prototyping of many sensor processing algorithms and database analytics.

Although the speed of this implementation is good, it is limited by the Galois Field matrix multiplication in the MATLAB Communications Toolbox. Other computational algebra systems have demonstrated extremely impressive speed on equivalent computations. For example, the Magma computational algebra system can perform the same Galois Field matrix multiplication 50x faster than MATLAB [Sutherland 2015]. Thus, additional optimization of the MATLAB Communications Toolbox could allow this implementation to equal or even surpass the speed of the OpenSSL version.

Other limitations in the MATLAB Communications Toolbox limit the ability to easily implement more complex cryptographic operations. For example, the widely used Galois Counter Mode (GCM) is difficult to implement without using 128 bit Galois Fields. Currently the MATLAB Communications Toolbox only supports 16 bit Galois Fields. As the MATLAB Communications Toolbox expands its support for Galois Fields, these more complex cryptographic modes should be easy to implement.

APPENDIX A1: MATLAB AES 128 BIT CBC ENCRYPTION IMPLEMENTATION

```
function ct = AESCBCencrypt(key,IV,pt)
  Nblock = ceil(size(pt,2)/16);                         % Number of 16 byte blocks for plaintext.
  pt(:,size(pt,2)+1:Nblock*16) = 0;                     % Pad plaintext to end of the last block.
  ct = uint8(pt);  ct(:) = 0;                           % Allocate ciphertext.
  Nround = 10;                                          % Standard for 128 bit.
  ob_gf8 = repmat(AESfield(uint8(IV)), size(pt,1), 1);  % Set up IV.
  sbox = genSbox();                                     % Compute sbox from first principles.
  row = [2 3 1 1];                                      % Define mix matrix.
  mixmat_gf8 = gallery('circul', row);
  key_schedule = genKeys(key, Nround, sbox);            % Generate the key schedule.
  idx = 1:16;
  for iblock=1:Nblock
    pt_gf8 = AESfield(pt(:,idx)) + ob_gf8;              % XOR plaintext with the IV.
    ctrnd00_gf8 = pt_gf8 + repmat(key_schedule(1,:), size(pt_gf8,1),1);    % Add plaintext and key.
    for iround=1:Nround
      ctrnd01_gf8 = AESfield(sbox(uint32(ctrnd00_gf8)+1));        % SubBytes
      ctrnd01_gf8(:,[2 3 4  6 7 8  10 11 12  14 15 16]) = ctrnd01_gf8(:,[6 11 16  10 15 4  14 3 8  2 7 12]);          % ShiftRows
      if (iround < Nround)                              % MixColumns
        ctrnd01_gf8(:,  1:4  ) = (mixmat_gf8 * ctrnd01_gf8(:,  1:4  ).').';
        ctrnd01_gf8(:,  5:8  ) = (mixmat_gf8 * ctrnd01_gf8(:,  5:8  ).').';
        ctrnd01_gf8(:,  9:12) = (mixmat_gf8 * ctrnd01_gf8(:,  9:12).').';
        ctrnd01_gf8(:,13:16) = (mixmat_gf8 * ctrnd01_gf8(:,13:16).').';
      end
      ctrnd01_gf8 = ctrnd01_gf8 + repmat(key_schedule(iround+1,:),size(pt_gf8,1),1);          % AddRoundKey
      ctrnd00_gf8 = ctrnd01_gf8;                        % Copy 01 variable to 00 variable for CBC mode.
    end
    ob_gf8 = ctrnd01_gf8;                               % Last ciphertext becomes the IV for the next block.
    ct(:,idx) = reshape(uint8(uint32(ctrnd01_gf8)), size(pt,1), []);  % Save current block of ciphertext.
    idx = idx + 16;                                     % Increment to the next block.
  end

function x = uint32(g);
  x = g.x;
end

function f = AESfield(inp)
  f = gf(inp,8,hex2dec('11B'));                         % Standard primitive polynomial for AES
end
```

APPENDIX A2: MATLAB AES 128 BIT CBC GENERATE S-BOX IMPLEMENTATION

```
function sbox = genSbox()
  pp = hex2dec('11B');                                  % primitive polynomial for AES .
  row = [1 1 1 1 1 0 0 0];                              % set up constants A and b.
  A = gf(gallery('circul', row), 2);
  b = gf([0; 1; 1; 0; 0; 0; 1; 1], 2);
  xinv = gf(0:255, 8, pp);                              % Compute inverses.
  xinv(2:256) = 1./xinv(2:256);
  y = gf(transpose(dec2bin(uint32(xinv),8) == '1'),2);
  z = A * y + repmat(b,1,256);                          % Affine transformation.
  sbox = transpose(uint8(bin2dec(num2str(transpose(uint32(z))))));        % Reformat to make sbox.
end
```

APPENDIX A3: MATLAB AES 128 BIT CBC GENERATE KEY SCHEDULE IMPLEMENTATION

```
function key_schedule = genKeys(key, Nround, sbox)
  if sbox == 0                                          % Encrypt can pass in the sbox, but decrypt uses the inverse sbox.
    sbox = genSbox();
  end
  key_schedule = AESfield(zeros(Nround+1,16));          % Allocate space for Nround+1 keys of size 4x4.
  key_schedule(1,:) = AESfield(uint8(key));             % Embed key into Galois Field.
  x = AESfield(2);                                      % Create round constant for any round.
  rcon = AESfield(zeros(1,4));
```

```matlab
  for iround=1:Nround
    rcon(1) = x^(iround - 1);                                           % Compute round constant.
    key_schedule(iround+1,1:4) = circshift(key_schedule(iround,13:16),-1,2);    % Shift column 4 and place into column 1.
    key_schedule(iround+1,:) = sbox(uint8(uint32(key_schedule(iround+1,:)))+1); % Tranform values with S-Box.
    key_schedule(iround+1,1:4) = key_schedule(iround+1,1:4) + rcon;     % Add round constant.
    % Add and shift forward to remaining columns.
    key_schedule(iround+1, 1:4 ) = key_schedule(iround, 1:4 ) + key_schedule(iround+1,1:4 );
    key_schedule(iround+1, 5:8 ) = key_schedule(iround, 5:8 ) + key_schedule(iround+1,1:4 );
    key_schedule(iround+1, 9:12) = key_schedule(iround, 9:12) + key_schedule(iround+1,5:8 );
    key_schedule(iround+1,13:16) = key_schedule(iround,13:16) + key_schedule(iround+1,9:12);
  end
end
```

APPENDIX A4: MATLAB AES 128 BIT CBC DECRYPTION IMPLEMENTATION

```matlab
function pt = AESCBCdecrypt(key,IV,ct)
  Nblock = size(ct,2)/16;                           % Number of 16 byte blocks in ciphertext.
  pt = char(zeros(size(ct,1), 16*Nblock));          % Preallocate for the plaintext.
  Nround = 10;                                      % Standard for 128 bit key.
  ob_gf8 = repmat(AESfield(uint8(IV)), size(ct,1), 1);  % Set up IV.
  sbox = genSboxInv();                              % Compute inverse sbox from first principles.
  row = [14 11 13 9];                               % Define mix matrix.
  mixmat_gf8 = gallery('circul', row);
  key_schedule = genKeys(key, Nround, 0);           % Generate the key schedule.
  idx = 1:16;
  for iblock=1:Nblock
  state00 = uint8(ct(:,idx));
  for iround=Nround:-1:1                            % Intermediate rounds.
    state01 = state00 + repmat(key_schedule(iround+1,:), size(ct,1),1);    % AddRoundKey
    if (iround < Nround)                            % MixColumns
      state01(:, 1:4 ) = (mixmat_gf8 * state01(:, 1:4 ).').';
      state01(:, 5:8 ) = (mixmat_gf8 * state01(:, 5:8 ).').';
      state01(:, 9:12) = (mixmat_gf8 * state01(:, 9:12).').';
      state01(:,13:16) = (mixmat_gf8 * state01(:,13:16).').';
    end
    state01(:, [2 3 4  6 7 8  10 11 12  14 15 16]) = state01(:, [14 11 8  2 15 12 6 3 16  10 7 4]);         % ShiftRows
    state01 = AESfield(sbox(uint32(state01)+1));    % SubBytes
    state00 = state01;                              % Copy 01 variable to 00 variable for CBC mode.
  end
  state01 = state01 + repmat(key_schedule(1,:),size(ct,1),1);    % Reverse round 0 (initial round).  Add plaintext and key.
  state01 = state01 + ob_gf8;                       % XOR state with the IV.
  ob_gf8 = uint8(ct(:,idx));                        % This ciphertext becomes the IV for the next block.
  pt(:,idx) = uint32(state01);                      % Save current block of plaintext.
  idx = idx + 16;                                   % Increment to the next block.
  end
end
```

APPENDIX A5: MATLAB AES 128 BIT CBC GENERATE INVERSE S-BOX IMPLEMENTATION

```matlab
function sinv = genSboxInv()
  pp = hex2dec('11B');                              % primitive polynomial for AES.
  row = [0 1 0 1 0 0 1 0];                          % set up constants A and b.
  A = gf(gallery('circul', row), 2);
  b = gf([0; 0; 0; 0; 0; 1; 0; 1], 2);
  y = gf(transpose(dec2bin(0:255) == '1'), 2);      % Affine transformation.
  z = A * y + repmat(b,1,256);
  z = gf(bin2dec(num2str(transpose(uint32(z)))), 8, pp);      % Reformat.
  zinv = gf(0:255, 8, pp);                          % Invert results.
  zinv(1:99) = 1./z(1:99);
  zinv(100) = 0;
  zinv(101:256) = 1./z(101:256);
  sinv = uint8(uint32(zinv));                       % Reformat to make inverse sbox.
end
```